# Anomaly in anomalous Nernst effect at low temperature for $C1_b$-type NiMnSb half-Heusler alloy thin film

Himanshu Sharma[1,2*], Zhenchao Wen[1,3], Koki Takanashi[1,3], Masaki Mizuguchi[1,2,3*]

[1]Institute for Materials Research, Tohoku University, Sendai, 980-8577, Japan
[2]CREST, Japan Science and Technology Agency, Kawaguchi 332-0012, Japan
[3]Center for Spintronics Research Network (CSRN), Tohoku University, Sendai, 980-8577, Japan
*E-mail: himsharma@imr.tohoku.ac.jp; mizuguchi@imr.tohoku.ac.jp

The anomaly in the anomalous Nernst effect (ANE) was observed for a $C1_b$-type NiMnSb half-Heusler alloy thin film deposited on a MgO (001) substrate. The Nernst angle ($\theta_{ANE}$) showed maximum peak with decreasing temperature and reached 0.15 at 80 K, which is considered to be brought by the cross-over from half-metal to normal ferromagnet in NiMnSb at low temperature. This anomaly was also observed for the transport properties, that is, both the resistivity and the anomalous Hall resistivity in the same temperature range.

## 1. Introduction

The anomalous Nernst effect (ANE), thermally induced transversal voltage in the direction perpendicular to both the temperature gradient and the magnetization, is a complimentary probe for the spin-orbit coupling phenomena in the field of spin caloritoronics [1-10]. The interplay between the ANE, pure spin-current and spin-polarized current generated by a thermal gradient has attracted many researchers from the last two decades. The ANE can be utilized to investigate the spin-orbit coupling phenomena through the interaction among spin, charge and heat. On the other hand, the $C1_b$-type half-metallic Heusler alloys (see a crystal structure in Fig. 1(a)) are of intense research interest because of their exclusive properties due to the semiconducting behavior of the minority band with a gap at the Fermi level (see the electronic band structure near the Fermi level in Fig. 1(b)) and thus leading to 100% spin polarization of conduction electrons [11-

24]. This exceptional property may direct distinctive thermoelectric effects in half-metallic Heusler alloys, which augment the interest in the physical properties in the concerned materials.

In this paper, we present a thorough investigation of the Seebeck and the ANE with in-plane applied thermal gradient (see in Fig. 1(c)) at different temperatures for a $C1_b$-type NiMnSb half-Heusler alloy thin film. The observed anomaly in thermoelectric effects is discussed with the temperature dependence of the electrical resistivity and the Hall resistivity.

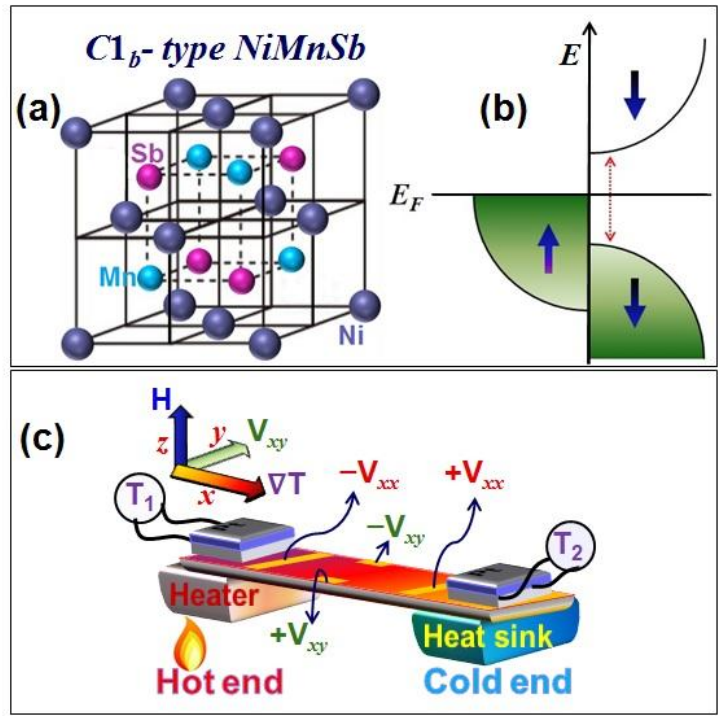

Fig. 1 (a) Schematic of the crystal structure with atom occupancy and (b) electronic band structure for $C1_b$-type NiMnSb half-Heusler alloy. (c) Schematic of device used to measure anomalous Nernst effect and Seebeck effect in NiMnSb thin film.

2. **Experimental**

A thin film of half-Heusler NiMnSb with a thickness of 20 nm was deposited by co-sputtering Ni and MnSb targets using DC-sputtering on a MgO (001) single crystalline substrate at room temperature. After deposition, the NiMnSb film was annealed at 300°C. The composition of the deposited NiMnSb thin film was evaluated to be $Ni_{1.01\pm0.02}Mn_{0.98\pm0.02}Sb_{1.01\pm0.02}$ by an inductively coupled plasma (ICP) analysis, which confirms the ideal stoichiometric composition for the half-Heusler compound [12].

In order to measure the ANE and the anomalous Hall effect (AHE), the NiMnSb thin film was patterned with Au electrodes using an optical lithography and ion-milling process. The typical lateral size of NiMnSb thin film channel was 2.2 mm × 5 mm, and the distance between two transverse voltage ($V_y$) probes and two longitudinal voltage ($V_x$) probes were 2 mm and 3 mm, respectively. A cryostat of a physical property measurement system was used to measure the ANE and AHE at low temperature and with high magnetic field. To create the in-plane thermal gradient across the structure along the x-axis, a ceramic heater at one (hot) end and a Cu-heat sink at the other (cold) end were used as shown in fig. 1(c) [6]. The pre-calibrated temperature sensors were used to sense the temperature difference across the structure, and the transverse output voltage ($V_y$) and longitudinal output voltage ($V_x$) were measured by a nanovoltmeter. The transverse Seebeck coefficient ($S_{xy}$) and the longitudinal Seebeck coefficient ($S_{xx}$) are calculated using the formulae, $[S_{xy} = E_y/\Delta T_x = \{+V_y - (-V_y)\}L/2w\Delta T_x]$ and $[S_{xx} = E_x/\Delta T_x = \{+V_x - (-V_x)\}/L\Delta T_x]$, respectively, where $\Delta T_x$ is the temperature difference applied along the x-axis to the sample and $L$ and $w$ are distance between the thermometry and width of sample [1-6].

## 3. Results and Discussion

The structural properties were characterized by X-ray diffraction (XRD) ($\theta$-$2\theta$ scan) pattern, as shown in Fig. 2. The diffraction peaks from NiMnSb (002) and (004) superlattices were clearly observed, which implies the chemical ordering of NiMnSb.

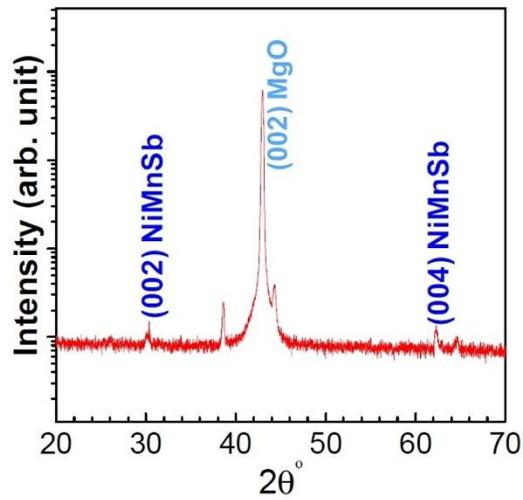

Fig. 2 XRD (θ-2θ-scan) pattern of a 20 nm thick NiMnSb thin film deposited on a MgO (001) substrate.

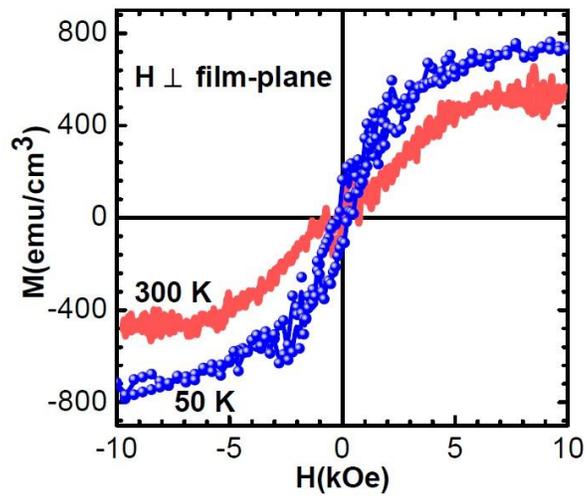

Fig. 3 The out-of-plane magnetization as a function of applied magnetic field ($H$) at 300 K and 50 K.

The out-of-plane magnetization as a function of applied magnetic field was measured at 50 K and 300 K using a superconducting quantum interference device as shown in Fig. 3. From the magnetization curve, the saturation magnetization ($M_s$) and the saturation magnetic field ($H_s$) are found to be approximately 500 emu/cm$^3$ and 6 kOe, respectively at 300 K. Whereas, at 50 K, the $M_s$ and $H_s$ increased significantly to be around 750 emu/cm$^3$ and 7.5 kOe, respectively.

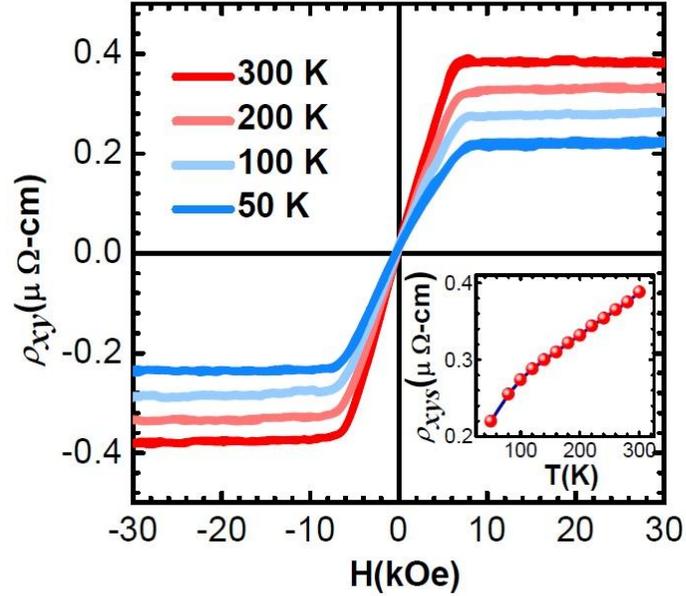

Fig. 4 Magnetic field dependence of the Hall resistivity ($\rho_{xy}$) measured at different temperatures with in-plane applied current of 100 µA. (Inset) The temperature dependence of saturation Hall resistivity ($\rho_{xys}$).

The Hall resistivity ($\rho_{xy}$) as a function of applied magnetic field ($H$) at different temperatures with in plane applied current of 100 µA was measured as shown in Fig. 4. The observed saturation magnetic field for $\rho_{xy}$ is same as that observed in magnetization data (see Fig. 3). Further, the temperature dependence of the saturation Hall resistivity ($\rho_{xys}$) is plotted in the inset of Fig. 4. It is observed from Fig. 4 that the $\rho_{xy}$ decreases linearly with decreasing temperature up to 120 K. However, below 120 K it further decreases with decreasing temperature with comparatively faster decreasing rate.

Figure 5 shows the magnetic field dependence of the transverse Seebeck coefficient ($S_{xy}$)

measured at different temperatures, when the applied magnetic field is perpendicular to the x-y plane, for the temperature difference of 3 K applied along the x-axis to the sample. The observed saturation magnetic field for $S_{xy}$ is the same as that observed in magnetization data and Hall data (see Fig. 3 and Fig. 4). A clear temperature dependent behavior of Nernst signal is observed in Fig. 5.

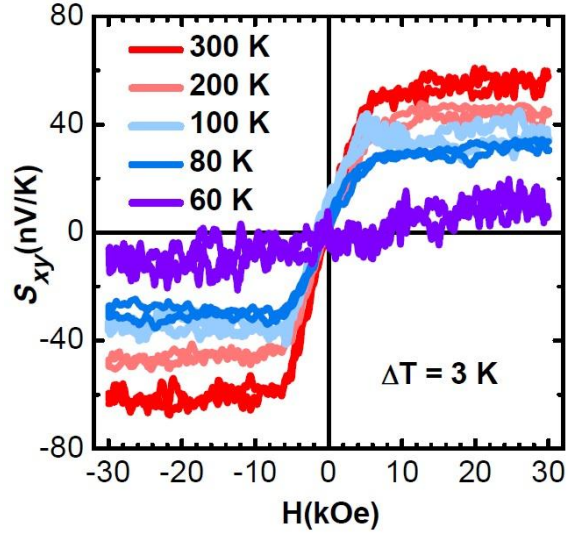

Fig. 5 Magnetic field dependence of the transverse Seebeck coefficient measured at different temperatures with in-plane applied temperature difference of 3 K.

In ferromagnets, the electromotive force ($E$) induced by spin-polarized current through ANE can be described as:

$$E = - S_{xy} \times \nabla T_x = \theta_{ANE} S_{xx} \times \nabla T_x \quad \text{-----（1）}$$

where $S_{xy}$ is the transverse Seebeck coefficient, $S_{xx}$ is the (longitudinal) Seebeck coefficient, and $\theta_{ANE}$ is the anomalous Nernst angle [1-5].

Figure 6(a) shows the temperature dependence of $S_{xy}$ (left axis) and $S_{xx}$ (right axis). It is observed that $S_{xy}$ simply decreases with decreasing temperature up to 120 K, and drops faster below 120 K. This temperature dependence behavior of $S_{xy}$ is consistent with that of $\rho_{xys}$ as plotted in the inset of Fig. 4.

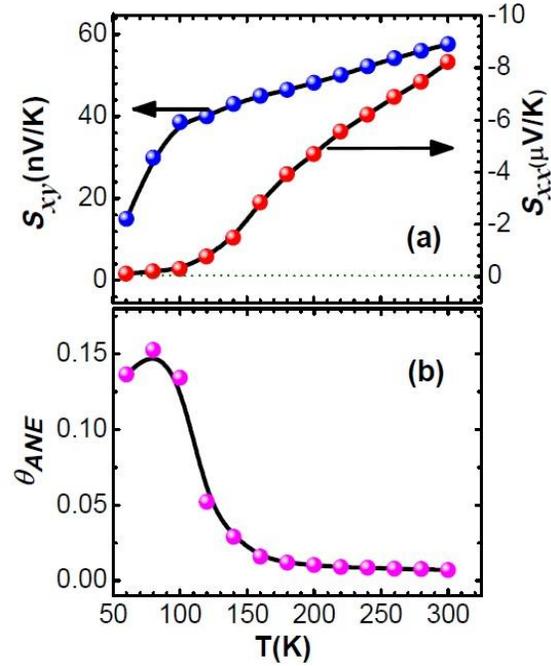

Fig. 6 (a) Temperature dependence of transverse Seebeck coefficient ($S_{xy}$) (left axis) and (longitudinal) Seebeck coefficient ($S_{xx}$) (right axis), (b) temperature dependence of Nernst angle ($\theta_{ANE}$).

The temperature dependence of $\theta_{ANE}$ is shown in Fig 6(b) as calculated from Eq. (1). It is observed that $\theta_{ANE}$ increases with decreasing temperature. More importantly, the $\theta_{ANE}$ below 100 K is found to be significantly enhanced and become about 10 times larger than that of 300 K. It is thought that this enhancement was brought by the anomaly in the temperature dependence of $S_{xx}$ at low temperature.

This distinct behavior observed in $\rho_{xy}$, $S_{xy}$ and $\theta_{ANE}$ at low temperature is probably associated with particular properties in NiMnSb leading to a low temperature anomaly [21,22]. The distinct mechanism of this anomaly observed in various physical properties has not clarified yet, whereas there are some reports which observed low temperature anomalies in various physical properties of NiMnSb single-crystal (bulk) and thin films [21,22]. They observed a cross-over from half-metal to normal ferromagnet in NiMnSb at low temperature in the temperature dependence of resistivity. $\rho(T)$ was proportional to $T^\alpha$ at low temperature with $1.7 < \alpha < 2.2$, while with $1.3 < \alpha < 1.5$ above 100 K. To confirm this behavior further in our sample, we have measured the temperature dependence of resistivity as shown in Fig. 7. In our resistivity data, we observed almost linear behavior of resistivity ($\rho_{xx}$) above 100 K and an irregular behavior of $\rho_{xx}$ below 100 K. Further, the derivative of $\rho_{xx}$ to highlight the irregular behavior of $\rho_{xx}$ below 100 K is plotted in inset of Fig. 7.

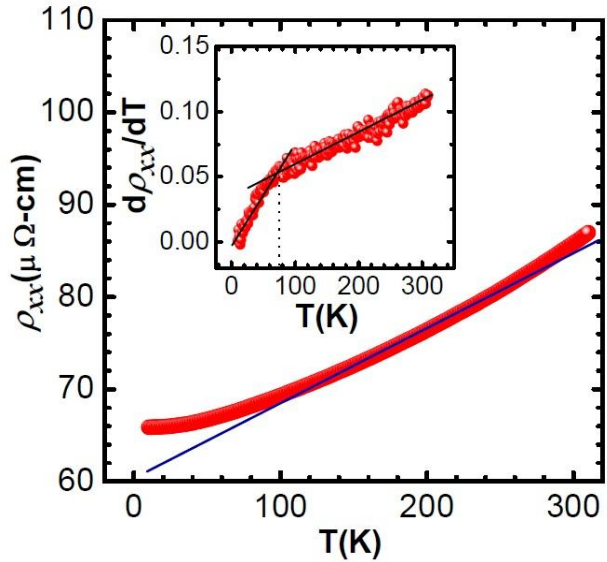

Fig. 7 Resistivity ($\rho_{xx}$) as a function of temperature and solid line shows the liner fitting. (Inset) The derivative of $\rho_{xx}$ to highlight the irregular behavior of $\rho_{xx}$ below 100 K. Solid lines show the liner fitting.

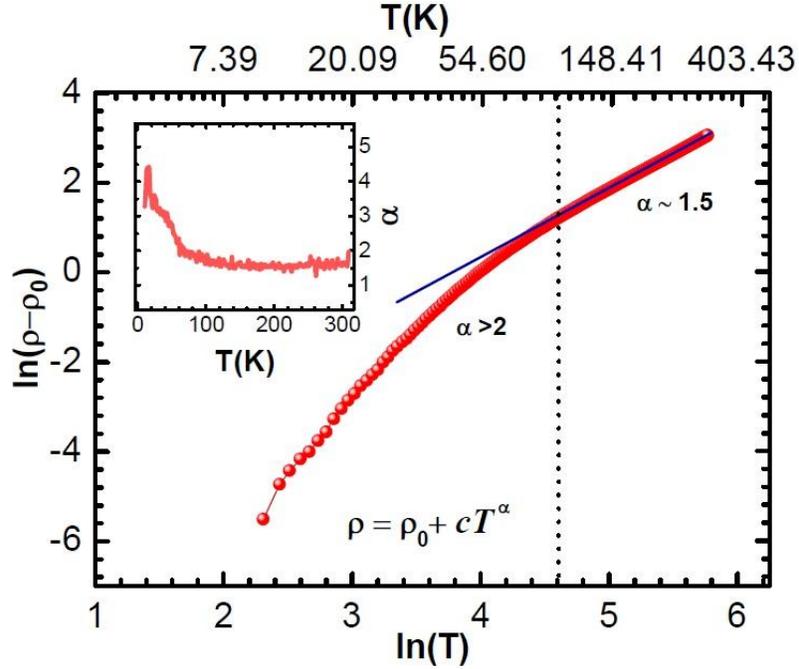

Fig. 8 Plot of ln(ρ – ρ$_0$) as a function of ln(T) by assuming the function of ρ = ρ$_0$ + C T$^α$ for our resistivity data. Line shows the liner fitting above 100 K. (Inset) The slope of curve as a function of T, which provides the local power-law factor α.

Next, in order to analyze the resistivity data of NiMnSb thin film, we have plotted ln(ρ – ρ$_0$) as a function of ln(T) as shown in Fig. 8. The slope of this curve corresponds to the power-law parameter "α", when the function of ρ = ρ$_0$ + C T$^α$, (where ρ$_0$ is the resistivity at 0 K) is assumed. The linear behavior of ρ for T > 100 K is shown by a solid line. The inset shows the variation of α as a function of temperature. In the low temperature region (T < 100 K), the resistivity behaves with α > 2, while with α is around 1.5 above 100 K. This result is consistent with that observed by previous reports [21-25]. Furthermore, in the critically low temperature region T < 40 K, we observed that the resistivity follows α > 3, which characterizes the Bloch-Gruneisen formula [21], and indicates a possibility of phonon scattering in this temperature range. In addition, the ANE in the NiMnSb thin film mainly originates from the thermally driven anomalous Hall conductivity, since the longitudinal electron transport is suppressed at low temperature, which can be confirmed by the consistent behavior of $\rho_{xys}$ (see in the inset of Fig. 4) and $S_{xy}$ (see in Fig. 6(a)).

We have provided further confirmation of anomaly observed at low temperatures in thermoelectric properties of NiMnSb thin film for the first time. Hence, this study will be helpful not only for exploring the physics of spin-orbit coupling phenomena using the interplay among heat, spin, pure spin-current and charge, but also help to explore the origin of anomaly observed at low temperature in NiMnSb. It will be interesting to see whether such anomaly observed in thermal and transport properties can also affect the anisotropy [26-28] of NiMnSb thin film.

## 4. Conclusions

In summary, the temperature dependence of the ANE and the Seebeck effects in addition to the resistivity and the anomalous Hall resistivity was investigated for a NiMnSb half-Heusler alloy thin film grown on a MgO substrate. Clear anomaly was observed below 100 K in not only the electrical transport but also in the ANE. The Nernst angle ($\theta_{ANE}$) showed maximum peak with decreasing temperature and reached 0.15 at 80 K, which is considered to be brought by the cross-over from half-metal to normal ferromagnet in NiMnSb at low temperature. We think this study will be helpful not only for exploring the physics of spin-orbit coupling phenomena using the interplay among heat, spin, pure spin-current and charge, but also help to explore the origin of anomaly observed at low temperature in NiMnSb.

**Acknowledgements**

This research was partly supported by JST-CREST (Grant No. JPMJCR1524), a Grant-in-Aid for Scientific Research (S) (25220910), Scientific Research (A) (17H01052), and Young Scientists B (17K14652) from Japan Society for the Promotion of Science, the Center for Spintronics Research Network (CSRN), and Collaborative Research Center on Energy Materials (E-IMR).